\documentclass[journal]{IEEEtran}
\IEEEoverridecommandlockouts
\usepackage[switch]{lineno}
\usepackage[nospace,noadjust]{cite}
\usepackage[english]{babel}
\usepackage{setspace} 
\usepackage[T1]{fontenc}
\usepackage{multirow}
\usepackage{amsmath,amssymb,amsfonts}
\usepackage{multicol}
\usepackage{multirow}
\usepackage{graphicx}
\usepackage{subfigure}
\usepackage{color}
\usepackage{textcomp}
\usepackage{soul}
\graphicspath{{images/}}
\usepackage[ruled,vlined]{algorithm2e}
\usepackage{tagging}

\usepackage{array}
\newcolumntype{L}{>{\centering\arraybackslash}m{3cm}}

\sethlcolor{yellow}
\usepackage[colorinlistoftodos,prependcaption,textsize=normalsize]{todonotes}
\usepackage{xargs}

\newcommandx{\doubt}[2][1=]{\todo[linecolor=red,backgroundcolor=red!25,bordercolor=red,#1]{#2}}

\usepackage[colorlinks=true, allcolors=blue]{hyperref}
\def\BibTeX{{\rm B\kern-.05em{\sc i\kern-.025em b}\kern-.08em
        T\kern-.1667em\lower.7ex\hbox{E}\kern-.125emX}}
\newcommand*{\email}[1]{%
    \normalsize\href{mailto:#1}{#1}\par
    }

\begin{document}
    \title{Automatic Crater Shape Retrieval using Unsupervised and Semi-Supervised Systems}
    \author{
    \IEEEauthorblockN{Atal Tewari\IEEEauthorrefmark{2}, Vikrant Jain\IEEEauthorrefmark{3}, Nitin Khanna\IEEEauthorrefmark{4}$^{,\star}
    $\thanks{$^{\star}$ Corresponding author. Please address all correspondences to Nitin Khanna, Multimedia Analysis and Security (MANAS) Lab, Electrical Engineering and Computer Science, Indian Institute of Technology Bhilai, India. E-mail address: \email{nitin@iitbhilai.ac.in}}
    }\\
    \IEEEauthorblockA{\IEEEauthorrefmark{2}Electrical Engineering, Indian Institute of Technology Gandhinagar, India}\\
    \IEEEauthorblockA{\IEEEauthorrefmark{3} Earth Sciences, Indian Institute of Technology Gandhinagar, India}\\
    \IEEEauthorblockA{\IEEEauthorrefmark{4}Electrical Engineering and Computer Science, Indian Institute of Technology Bhilai, India}    
}

\maketitle  
\thispagestyle{plain}
\pagestyle{plain}

\begin{abstract}
Impact craters are formed due to continuous impacts on the surface of planetary bodies. 
Most recent deep learning-based crater detection methods treat craters as circular shapes, and less attention is paid to extracting the exact shapes of craters. 
Extracting precise shapes of the craters can be helpful for many advanced analyses, such as crater formation.
This paper proposes a combination of unsupervised non-deep learning and semi-supervised deep learning approach to accurately extract shapes of the craters and detect missing craters from the existing catalog. 
In unsupervised non-deep learning, we have proposed an adaptive rim extraction algorithm to extract craters' shapes. In this adaptive rim extraction algorithm,  we utilized the elevation profiles of DEMs and applied morphological operation on DEM-derived slopes to extract craters' shapes. The extracted shapes of the craters are used in semi-supervised deep learning to get the locations, size, and refined shapes. 
Further, the extracted shapes of the craters are utilized to improve the estimate of the craters' diameter, depth, and other morphological factors. The craters' shape, estimated diameter, and depth with other morphological factors will be publicly available. 

\end{abstract}


\section{Introduction}

Craters are one of the most abundant features on the lunar surface. Crater shape, size, and frequency distribution can be used for finding the age of the surface~\cite{michael2010planetary}, precise spacecraft landing~\cite{downes2020lunar}, and understanding the geological processes~\cite{wilhelms1987geologic}.
Manual extraction of craters and retrieving their shape is time-consuming and error-prone. To address this many automatic crater detection methods were proposed.
Recently, deep learning (DL) based crater detection methods (e.g.~\cite{yang2022progressive,fan2022elcd,yang2021craterdanet,silburt2019lunar,yang2021high,ali2020automated}) have significantly improved the performance of crater detection on larger regions of planetary surfaces. 
However, most deep learning-based crater detection methods treat craters as circular objects and very less attention is given to extracting exact shapes of craters.

Extracting exact shapes of the craters helps to understand the formation, mechanisms, and spatial heterogeneity of craters on the moon~\cite{liu2017boundary}. 
Further, morphological features such as depth and depth-diameter ratio can be estimated with greater accuracy. 
These morphological features can be used to understand many physical characteristics of lunar surface, such as depth allows us to estimate the thickness of lunar regolith~\cite{di2016lunar}.
The proposed system automatically extracts the shapes of the craters in addition to automatic crater detection. 

The main contributions of this paper are as follows: 

\begin{itemize}
    \item Proposed an unsupervised algorithm, i.e., adaptive rim extraction algorithm using elevation profile with the help of morphological operations on DEM-derived slope to extract the shapes of the craters. Further, it is refined using a deep learning-based approach, i.e., Mask R-CNN~\cite{he2017mask}.
    \item Improved estimates of diameter, depth, and other morphological factors of the craters are obtained. 
\end{itemize}

This paper describes the proposed methodology and detail of crater morphological factors in Section~\ref{sec:methods}. Section~\ref{sec:experimental results} contains the experimental results, which include lunar crater detection, shape retrieval performance, and transfer learning on the Martian surface. Finally, concluding remarks and future work are given in Section~\ref{sec:Conclusion and Future Work}.

\section{Methodology} 
\label{sec:methods}

\begin{figure*}[!ht]
	\centering
	\includegraphics[width=0.93\textwidth,trim=0 0cm 0cm 0, clip]{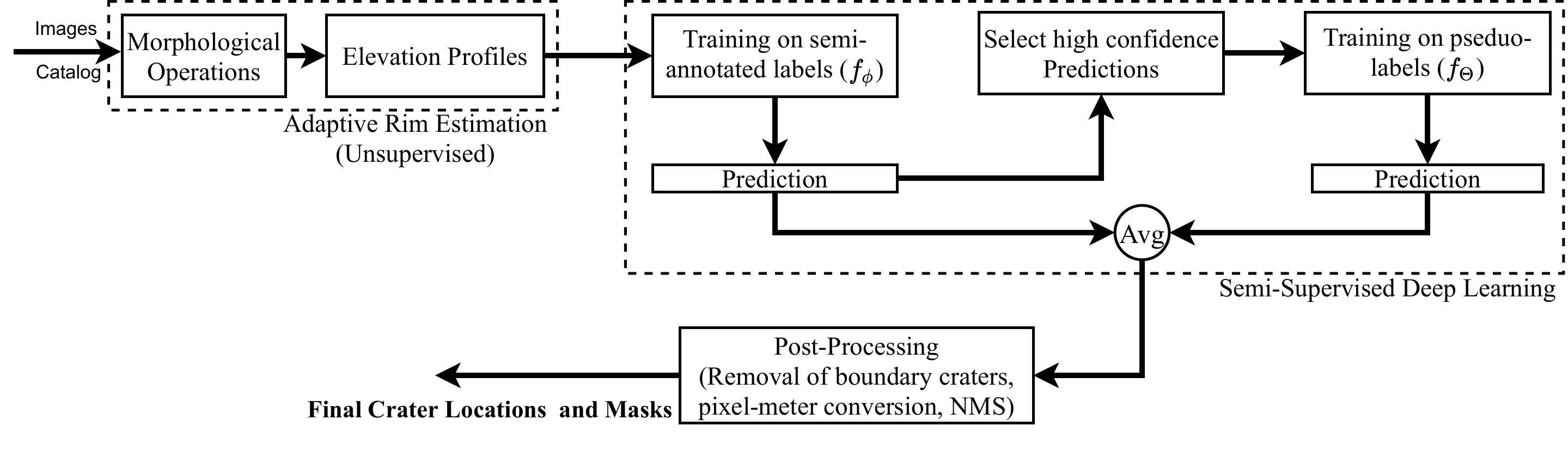}
	\caption{An Overview of the Proposed System.}
	\label{fig:oveview_of_our_work}
\end{figure*}

An overview of the proposed system is shown in Figure~\ref{fig:oveview_of_our_work}. 
The proposed system consists of two stages. 
First, we extract the crater's shape using adaptive rim extraction (Section~\ref{subsec:Adaptive Rim calculation using elevation profile}). 
Then the extracted rim is refined using the deep learning-based method, i.e., Mask R-CNN~\cite{he2017mask}. 
To detect craters missing from the ground truth, we further train a deep learning framework using high confidence predicted masks to detect the new craters.

\subsection{Dataset Preparation}
\label{subsec:Dataset Preparation}

Our system utilizes the lunar DEM dataset generated by Tewari et al.~\cite{TEWARI2022105500}. 
The DEM was generated using the lunar orbiter laser altimeter (LOLA) in the lunar reconnaissance orbiter (LRO) and SELENE terrain camera (TC)~\cite{bib:barker2016new}. 
The resolution and image size used is $100$ m/pixel and $512\times512$ pixels. 
However, in this work, some of the steps in the dataset generation process are modified. 
First, we transform the data from simple cylindrical to orthographic projection for reducing the projection distortion. 
Second, masks generated from the adaptive rim estimation algorithm are used in training the deep learning framework instead of the circular masks.

This study spans longitude -$180^\circ$ to $180^\circ$ and latitude $\pm60^\circ$. For training, longitude -$180^\circ$ to $60^\circ$ and latitude $\pm60^\circ$ is considered, and for testing,  longitude $60^\circ$ to $180^\circ$ and latitude $\pm60^\circ$ is considered. 
We used Povilaitis et al.~\cite{povilaitis2018crater} catalog for ground truth. It contains craters diameter size $5$ to $20$ km. 
The total craters in training and testing are $11559$ and $7764$.

\subsection{Adaptive Rim Estimation using Elevation Profile and Morphological Operations}
\label{subsec:Adaptive Rim calculation using elevation profile}

We have utilized the elevation profile of DEM for adaptive rim estimation. Similar to~\cite{agarwal2019study}, we are considering the highest elevation point as the rim point (boundary point of the crater).
From the known location of crater center ($x_c$,$y_c$) and radius ($r$) (Povilaitis et al. catalog~\cite{povilaitis2018crater}), in a direction $\theta$, we search for the maximum elevation value between the start point ($l_{start}$) = $0.5r$ and end point ($l_{end}$) = $r+10$. 
If the maximum elevation value occurs at a point between $l_{start}$ and $l_{end}$, this location corresponding to the maximum value will be considered as a rim point. 
If not, we shift the range by $l_{step}$ pixels. 
This process is repeated till the rim point is found or $l_{end} \geq 1.6r$.
Similarly, at every angle ($\theta$) with step size ($\theta_{step}$), rim points of the crater are estimated. 
Algorithm~\ref{algo:Adaptive Rim calculation using Elevation Profile} presents the details of this procedure. 
In our case, we fixed $l_{step}$ to $5$ and $\theta_{step}$ to $2^\circ$.

\begin{algorithm}[ht!]
\SetAlgoLined{
\textbf{Input:}
$I(x,y)$: DEM.\\
$\theta_{step}$: Finding rim location in every $\theta_{step}$ angle.\\
$l_{step}$: Moving step along elevation profile. 
\\
Center location ($x_c,y_c$) and radius ($r$) of craters in catalog. \\

\textbf{Output:}
Location of rim points  $(x^{rim}_{\theta}, y^{rim}_{\theta})$, for every $\theta = k\theta_{step}$, for integers $k$, $0 \leq k < 360/\theta_{step} $.\\

\textbf{Now}, for each crater in DEM, extract rim region from Algorithm~\ref{algo:Morphological Processing}.

\textbf{For} each angle $\theta = k\theta_{step}$, for integers $k$, $0 \leq k < 360/\theta_{step} $ \textbf{do}, \\
            \begin{enumerate}
                \item \textbf{If} along the ray starting from ($x_c,y_c$) at an angle $\theta$, no foreground pixel exists in the rim region between $0.3r$ and $1.6r$ from the center \textbf{then}, 
                \begin{enumerate}                
                    \item $x^{rim}_{\theta} = x_c + r \cos(\theta)$; $y^{rim}_{\theta} = y_c + r \sin(\theta)$
                \end{enumerate}            
                \item \textbf{else},
                start point ($l_{start}$) is $0.5r$ and end point ($l_{end}$) is $r + 10$ ;\newline
                \textbf{While} True \textbf{do}
                \begin{enumerate}                
                \item if $l_{end} \geq 1.6r$ then
                \begin{enumerate}
                    \item $x^{rim}_{\theta} = x_c + r \cos(\theta)$; $y^{rim}_{\theta} = y_c + r \sin(\theta)$
                    \item break
                \end{enumerate}                
                \item Calculate the elevation values,
                $E(l)$ ($E(l) = I(x_c + l\cos(\theta),y_c + l\sin(\theta)); l \in [l_{start},l_{end}]$);                
                \item If $\arg\max(E(l))\in [l_{start}, l_{end})$ then
                \begin{enumerate}
                    \item $x^{rim}_{\theta} = x_c + \arg\max(E(l)) \cos(\theta)$
                    \item $y^{rim}_{\theta} = y_c + \arg\max(E(l)) \sin(\theta)$
                    \item break
                \end{enumerate}
                \item else
                \begin{enumerate}
                    \item $l_{start} = l_{start} + l_{step}$; $l_{end} = l_{end} + l_{step}$
                \end{enumerate}
                \end{enumerate}                            
            \end{enumerate}            

\caption{Adaptive Rim Estimation using Elevation Profiles}}
\label{algo:Adaptive Rim calculation using Elevation Profile}
\end{algorithm}

If the rim point does not exist in a specific direction of a crater due to crater degradation, then using only the elevation profile to calculate the rim point will yield an incorrect rim point.
To tackle this issue, first of all the rim region is extracted by applying a set of morphological operations on the DEM-derived slope raster. 
The steps of morphological operations for finding the rim region are provided in Algorithm~\ref{algo:Morphological Processing}.
The output of these morphological operations is a binarized image, where the foreground pixels represent the potential rim points of a crater.
If a potential rim point exists in a direction, then the corresponding elevation profile is used to extract the rim point (Algorithm~\ref{algo:Adaptive Rim calculation using Elevation Profile}, step $2$); otherwise, the rim point is marked at the radius provided by the catalog. 

\begin{figure*}[!htb]
    \centering
    \includegraphics[width=0.95\textwidth,trim=0.4cm 0.18cm 0.7cm 0.85cm, clip]{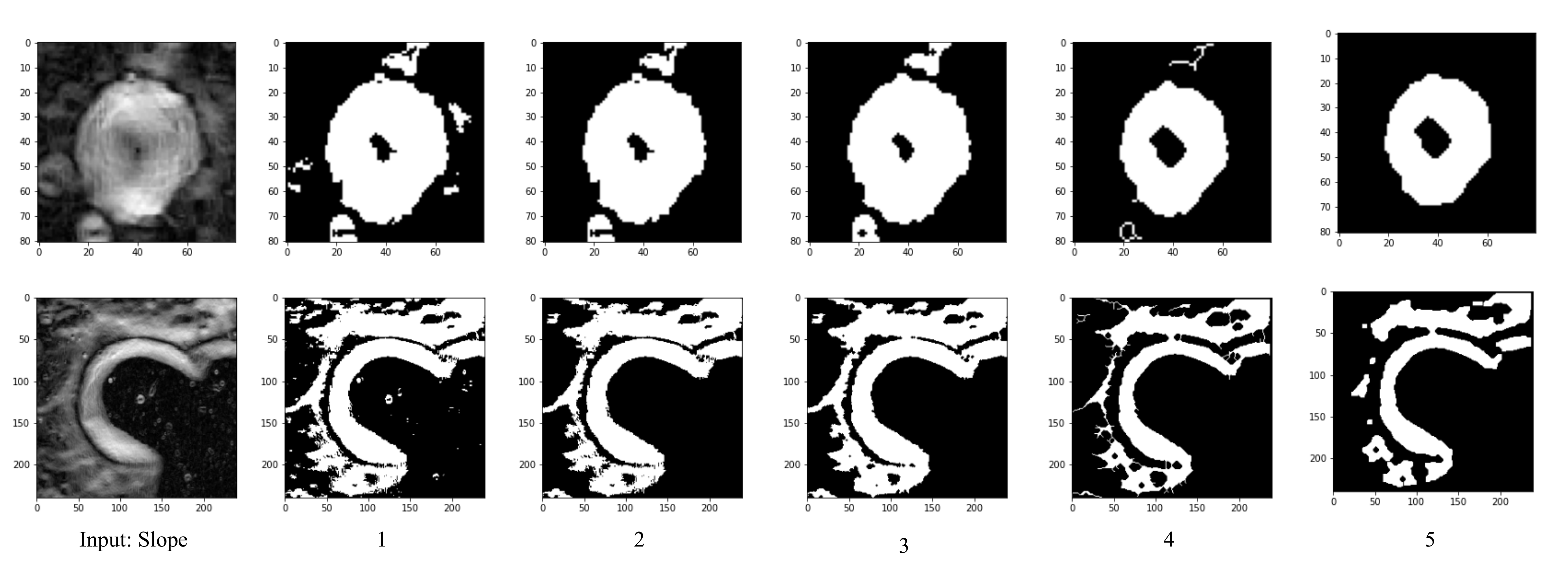}
    \caption{Visual inspection for Steps of Morphological Operation (Algorithm~\ref{algo:Morphological Processing}).}
    \label{fig:morphological_operation}
\end{figure*}

Figure~\ref{fig:morphological_operation} shows the visual results for a few craters to demonstrate how the morphological operation Algorithm~\ref{algo:Morphological Processing} works in each step. 
In this figure, foreground pixels (white regions) are potential rim regions. 
As shown in the figure, noisy regions are significantly reduced from step 1 to step 5. 
Also, Algorithm~\ref{algo:Morphological Processing} successfully distinguishes degraded and non-degraded regions (second row in Figure~\ref{fig:morphological_operation}). 
The output of Step 5 is used in Algorithm~\ref{algo:Adaptive Rim calculation using Elevation Profile} to avoid estimating the rim in the directions where foreground pixels do not exist. 

\begin{algorithm}[ht!]
\SetAlgoLined{
\textbf{Input:}
Slope raster, center position ($x_c$,$y_c$) and radius ($r$) of the crater

\textbf{Output:} 
Binary image corresponding to rim region where foreground pixels are potential rim points.
\newline  
\newline
For each crater region (top corner: ($x_c-1.6r$, $y_c-1.6r$), bottom corner: ($x_c+1.6r$, $y_c+1.6r$)) in slope  raster do:
\begin{enumerate}
    \item Binarize it using the Otsu's thresholding method;
    \item Remove small objects to reduce outliers and/or noise using connected component method;
    \item Perform closing operation to remove unwanted pixels and fill gaps around crater boundary;
    \item Perform thinning to reduce boundary region pixels;
    \item Finally, perform erosion followed by dilation operation to remove unwanted pixels.
\end{enumerate}
\caption{Morphological Operations}}
\label{algo:Morphological Processing}
\end{algorithm}

\subsection{Semi-supervised Deep Learning}
\label{subsec:Iterative Training on Mask R-CNN}
 
For training the network, we are using Povilaitis et al.~\cite{povilaitis2018crater} catalog, which is a conservative catalog. 
Thus, many craters are unlabelled in this catalog. 
Therefore, we follow a semi-supervised approach~\cite{lee2013pseudo}, which improves supervised learning using unlabeled data. The flow of semi-supervised deep learning is shown in Figure~\ref{fig:oveview_of_our_work}.
We first train the network, $f_\phi$, using crater shapes obtained from the proposed adaptive rim estimation algorithm. 
Then we predict the labels (shape, location, and size of the craters) using this network. 
This network ($f_\phi$) is able to predict many craters which are not labelled in the catalog. 
The high confidence predicted labels are used to retrain the network, $f_\Theta$. The predicted labels from $f_\phi$, which are used for training the network $f_\Theta$, are called pseudo-labels. 
In the proposed system, predicted labels with a confidence score $\geq 0.85$ are used for training the network $f_\Theta$. 
If only one model detects a crater ($f_\phi$ or $f_\Theta$), that model's predicted crater shape is considered as the average. 
Further, the craters' locations and sizes are calculated using these shapes. 

We have used a Mask R-CNN for training the networks $f_\phi$ and $f_\Theta$. 
Mask R-CNN is an instance segmentation method, an extension of the Faster R-CNN~\cite{ren2017faster} and adds a parallel branch for predicting the mask along with a bounding box.
Mask R-CNN consists of two stages. 
The first stage contains the feature extraction network and region proposal network (RPN) to extract the features from input and find the potential locations and sizes of the craters. 
Extracted region proposals (potential craters) from the first stage are passed to the second stage. The second stage utilizes these proposals to detect and segment the craters. 
Further details of Mask R-CNN are provided in the original paper~\cite{he2017mask}.

\subsection{Morphological Factors for Craters}
\label{subsect:Morphological Feature for Craters}

\begin{figure}[!ht]
    \centering
    \includegraphics[width=0.29\textwidth,trim=6cm 4.5cm 5cm 3cm, clip]{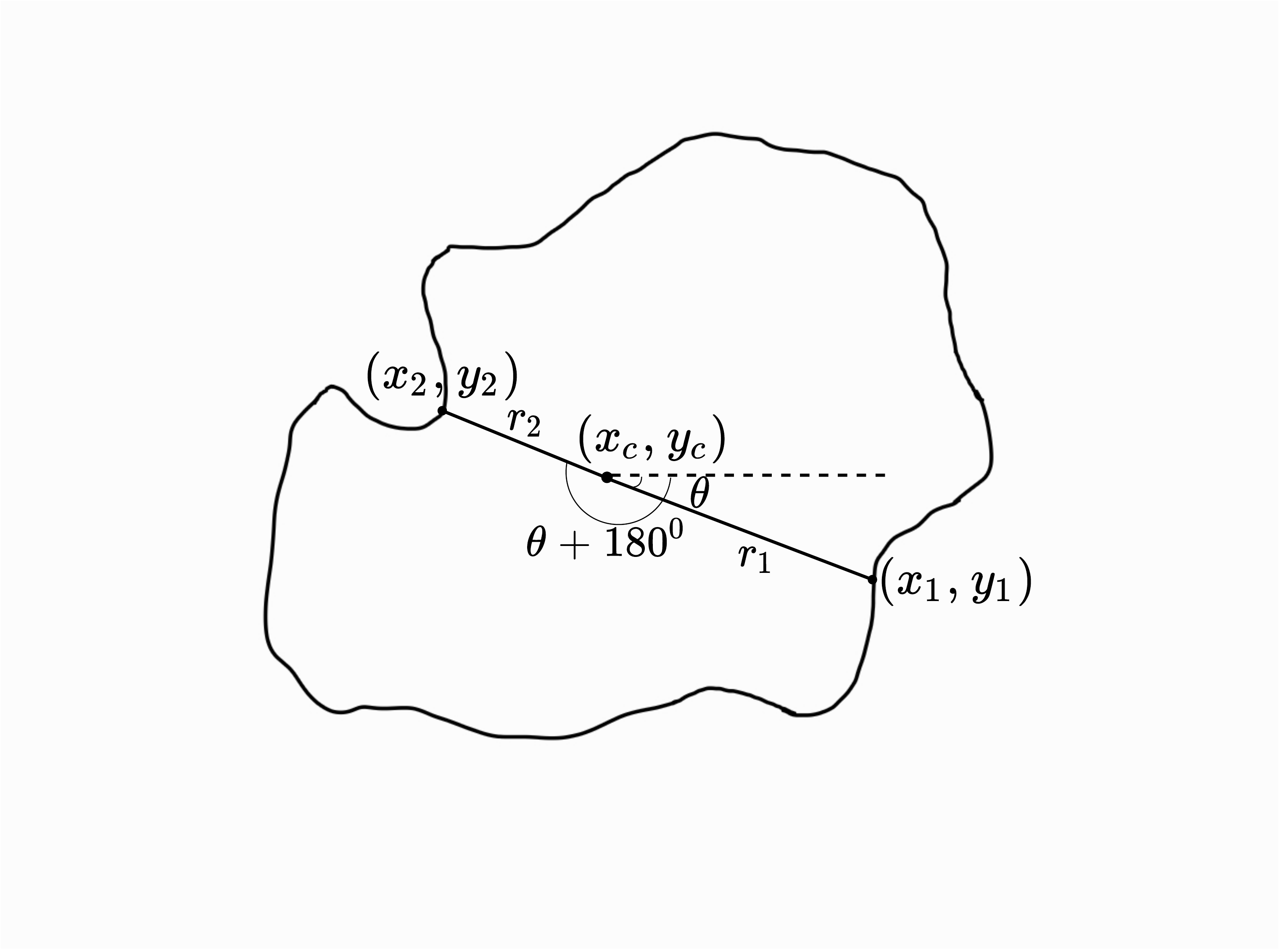}
    \caption{Schematic diagram of the crater for estimating diameter and depth at a particular rim point.}
    \label{fig:avg_depth_diameter_calculation}
\end{figure}

To calculate craters' morphological factors, we have utilized the extracted shapes from the proposed approach. 
The diameter and depth calculation at a particular rim point of the crater is shown in Figure~\ref{fig:avg_depth_diameter_calculation}. 
First, radius ($r_1$) and angle ($\theta$) are calculated using rim point $(x_1,y_1)$ (point A) and center point $(x_c,y_c)$. 
Then in direction, $\theta + 180^\circ$, $(x_2,y_2)$ (point B) is found and radius ($r_2$) is calculated. 
Finally, we add the radius $r_1$ and $r_2$ to get the diameter at a particular rim point.
Similarly, we calculated the diameter at all the rim points of the crater. 
The average of diameter values corresponding to all the rim points of a crater is used as the final estimate of a crater's diameter.
In addition to the average diameter, we have also provided each crater's minimum and maximum diameter.  

The crater depth varies in different directions; therefore, the average depth needs to be estimated for craters.   
Let $E(x,y)$ denotes the elevation value at location $(x,y)$ (Figure~\ref{fig:avg_depth_diameter_calculation}). 
The minimum elevation is calculated along the line from $A$ to $B$ (say $L_{AB}$). 
We used $E(x_1,y_1)$, $E(x_2,y_2)$, and minimum elevation value on $L_{AB}$ line (say $E(x_{min},y_{min})$) to estimate the depth of the crater at a particular rim point. 
Therefore, the depth value at $(x_1,y_1)$ is $E(x_1,y_1)$-$E(x_{min},y_{min})$, and depth value at $(x_2,y_2)$ is $E(x_2,y_2)$-$E(x_{min},y_{min})$. 
Similarly, depth values are calculated at other rim points. 
We have provided each crater's average depth, minimum depth, maximum depth, and the ratio (minimum depth)/(maximum depth). 
The ratio (minimum depth)/(maximum depth) can be used to understand the craters' degradation status. 
Further, estimated average diameter and depth are used to calculate the depth-diameter ratio.

\begin{table*}[!ht]
    \centering
    \caption{Description of morphological factors of the craters}\label{tab:morphological_index}
    \resizebox{0.95\textwidth}{!}
    {    
    \begin{tabular}{p{0.105\textwidth}p{0.605\textwidth}p{0.299\textwidth}}
    \hline
    Index     & Formula &  Significance\\
    \hline
    Circularity &  $\frac{4\pi A}{P^2}$, where, $A$ is the area and $P$ is the perimeter of the crater & Measure of the irregularity of the crater's rim\\
    Rectangle Factor & $\frac{A}{A_r}$, where, $A$ is the area of the impact crater, and $A_r$ is area of the minimum bounding rectangle & Measure of the fullness of the crater \\
    Sphericity & $\frac{R_i}{R_o}$, where, $R_i$ is radius of inscribed circle, $R_o$ is radius of ex-circle  & Measure of the closeness of crater as a circle\\
    Posture ration & $\frac{W}{L}$, where $W$ is the width of the minimum bounding rectangle and $L$ is the length of minimum bounding rectangle ($L \geq W$) & Measure of the narrowness of the crater\\
    Posture angle & Angle between the line along the length ($L$) of the minimum bounding rectangle of the crater and horizontal axis & Measure of the orientation of the impacts\\
    \hline
    \end{tabular}
    \label{tab:my_label}
    }
\end{table*}

Similar to~\cite{chen2017morphological,liu2017boundary,chen2018lunar}, we have also calculated other morphological factors of the craters, i.e., circularity, rectangular factor, sphericity, posture ratio (Table~\ref{tab:morphological_index}). 
In addition, we also calculated the angle between the line along the length of the minimum bounding rectangle of the crater and horizontal axis, which we called posture angle. 
The calculated morphological factors, along with craters' locations and sizes will be shared with the published version of the paper.

\subsection{Post-processing}
\label{subsec:Post-processing}

Similar to Tewari et al.~\cite{TEWARI2022105500}, in post-processing, we remove the boundary craters, convert the craters' locations from pixel coordinate to the geographic coordinate, and finally, used non-maximum suppression (NMS) to remove the duplicate craters. 
In this work, the steps followed for boundary crater removal and the NMS approach are the same as in~\cite{TEWARI2022105500}, whereas, for pixel coordinate to meter coordinate conversion, we have followed corresponding steps in Silburt et al.~\cite{silburt2019lunar}.

\subsection{Performance Evaluation Metrics}
\label{subsec:Performance Evaluation Metrics}

Crater detection performance is evaluated using precision, recall, F$_1$-Score, and F$_2$-Score.
These are calculated as,
\begin{align*}
\text{Precision} &= \frac{TP}{TP + FP}\\
\text{Recall} &= \frac{TP}{TP + FN}\\
\text{F$_1$-Score} &= \frac{2 \times \text{Precision} \times \text{Recall}}{\text{Precision} + \text{Recall}}\\
\text{F$_2$-Score} &= \frac{5 \times \text{Precision} \times \text{Recall}}{4 \times \text{Precision} + \text{Recall}}    
\end{align*}
Where, $TP$ is true positive, i.e., the number of detected craters present in the ground truth. $FP$ is false positive, i.e., the number of detected craters that are not present in the ground truth. $F_1$-Score balances the precision and recall values. 
$F_2$-Score emphasizes the recall over precision. 

\section{Experimental Results}
\label{sec:experimental results}
In this paper, we have utilized Mask R-CNN implementation~\cite{matterport_maskrcnn_2017} in Keras with TensorFlow backend. To increase the variability of crater features following augmentation is done. 
The image is rotated by $90^\circ$, $180^\circ$ and $270^\circ$, brightness values are changed by $80\%$ to $150\%$ from original values, Gaussian blur is applied with sigma value $0$ to $5$, and flipping is done horizontally and vertically from the center. 
In our experiments, learning rate is 0.01, SGD optimizer, epochs $20$, and image size feed to the network is $512 \times 512$ pixels.

\subsection{Finding optimal post-processing parameters}
\label{subsec:Finding optimal post-processing parameters}

We have analyzed the effect of post processing parameters, i.e., boundary threshold (m) and NMS threshold ($\delta$). The best parameter is chosen based on best $F_2$-Score in validation set. The $m$ $\in$ $\{0,1,5,10,15,20,25\}$ and $\delta$ $\in$ $\{0.1,0.2,0.3,0.4,0.5\}$ values are optimized simultaneously. 

The experiment of removal of boundary craters is shown in Table~\ref{tab:Rmv_bdry}. It can be seen that increasing the value of $m$ increases the precision value and decreases the recall value. The best $F_2$-score is at $m$$=$$15$. 

\begin{table}[ht!]
	\centering
	\caption{Effect of Removing Boundary Craters ($\delta = 0.4$) }\label{tab:Rmv_bdry}
	\begin{tabular}{p{0.02\textwidth} p{0.09\textwidth} p{0.08\textwidth}p{0.09\textwidth}p{0.09\textwidth}} 
		\hline
		$m$ & Precision (\%) & Recall (\%) & F$_1$-score (\%) & F$_2$-score (\%)\\ [0.5ex] 
		\hline
	    0 & 52.68 & 95.82 & 67.99 & 82.33 \\ 
		1 & 54.70 & 95.48 & 69.55 & 83.09 \\ 

		5 & 56.83 & 94.08 & 70.86 & 83.18 \\

		10 & 57.52 & 93.70 & 71.29 & 83.23 \\
		15 & 58.24 & 93.45 & 71.76 & 83.37 \\
		20 & 58.61 & 92.94 & 71.89 & 83.19 \\
		25 & 59.01 & 92.69 & 72.11 & 83.19 \\
		\hline
	\end{tabular}
\end{table}

Table~\ref{tab:Head-LROC_NMS} shows the effect of NMS on performance. It can be seen that with using NMS (except the last row) and without NMS (last row), the performance is significantly improved. 
Further, we can see that increasing the NMS threshold parameter ($\delta$) decreases the precision and increases the recall value. The best $F_2$-score is at $\delta$=$0.4$. 

Based on the experiments of removing boundary craters (Table~\ref{tab:Rmv_bdry}) and NMS (Table~\ref{tab:Head-LROC_NMS}), we got the optimal $F_2$-score at $m=15$ and $\delta=0.4$, which are used for further experiments.

\begin{table}[ht!]
	\centering
	\caption{Effect of using NMS ($m = 15$)}\label{tab:Head-LROC_NMS}
	\begin{tabular}{p{0.07\textwidth}p{0.065\textwidth} p{0.065\textwidth}p{0.075\textwidth}p{0.075\textwidth}}
		\hline
		$\delta$ & Precision (\%) & Recall (\%) & F$_1$-score (\%) & F$_2$-score (\%) \\ [0.5ex] 
		\hline
		0.1 & 58.69 & 91.93 & 71.64 & 82.57 \\ 
		0.2 & 58.65 & 92.81 & 71.88 & 83.13\\
		0.3 & 58.51 & 93.11 & 71.86 & 83.26\\
		0.4 & 58.24 & 93.45 & 71.76 & 83.37 \\
		0.5 & 57.59  & 93.62 & 71.31 & 83.21 \\
		Without NMS & 23.31 & 94.12 & 37.37 & 58.56 \\
		\hline
	\end{tabular}
\end{table}

\subsection{Crater Detection and Shape Retrieval Performance}
\label{subsec:Crater Detection Performance}

The crater detection performance of the proposed system is compared with DeepMoon~\cite{silburt2019lunar}, DeepCraters~\cite{yang2020lunar}, and Tewari et al.~\cite{TEWARI2022105500}. 
We have chosen these systems for comparison as the craters' location and size information provided by these papers is in the same region (longitude: $60^{\circ}$ to $180^{\circ}$ and latitude $\pm 60^{\circ}$) and diameter range ($5-20$ km) as we have considered for testing. 
The crater location  and size information of DeepMoon~\cite{silburt2019lunar}, DeepCraters~\cite{yang2020lunar}, and Ali-Dib et al.~\cite{ali2020automated} are utilized from following links~\footnote{\url{https://doi.org/10.5281/zenodo.1133969}}$^,$\footnote{\url{https://doi.org/10.6084/m9.figshare.12768539.v1}}$^,$\footnote{\url{https://github.com/malidib/Craters_MaskRCNN/}}.
Tewari et al.~\cite{TEWARI2022105500} results are directly taken from their paper.

\begin{table}[ht!]
	\centering
	\caption{Comparison with Existing Systems.}\label{tab:Crater detection performance}
	\begin{tabular}
	{p{0.14\textwidth} p{0.06\textwidth} p{0.06\textwidth}p{0.07\textwidth}p{0.07\textwidth}} 

		\hline
		Model Name & Precision (\%) & Recall (\%) & F$_1$-score (\%) & F$_2$-score (\%) \\ [0.5ex]
		\hline

		Proposed & 50.06 & \textbf{96.86}  & 66.00 & 81.60 \\ 

		Tewari et al.~\cite{TEWARI2022105500} & 57.70 & 95.71  & 71.99 & \textbf{84.57}  \\ 
		DeepCraters~\cite{yang2020lunar}  & 42.15 & 75.68 & 54.15 & 65.29	\\			
		DeepMoon~\cite{silburt2019lunar}  & \textbf{61.74} & 87.42 & \textbf{72.37} & 80.71  \\
		\hline
	\end{tabular}
\end{table}

Table~\ref{tab:Crater detection performance} shows the crater detection performance of the proposed and existing systems. 
The recall value of the proposed system is higher compared to existing systems. 
However, the precision value is lower compared to~\cite{silburt2019lunar,TEWARI2022105500}. 
As reasoned in Silburt et al.~\cite{silburt2019lunar}, the lower precision can be due to the newly detected craters because we have utilized a conservative catalog~\cite{povilaitis2018crater} that has only highly certain craters. 

\begin{figure*}[!ht]
	\centering
	\includegraphics[width=0.97\textwidth,trim=0.16cm 0.25cm 0.58cm 0.25cm, clip]{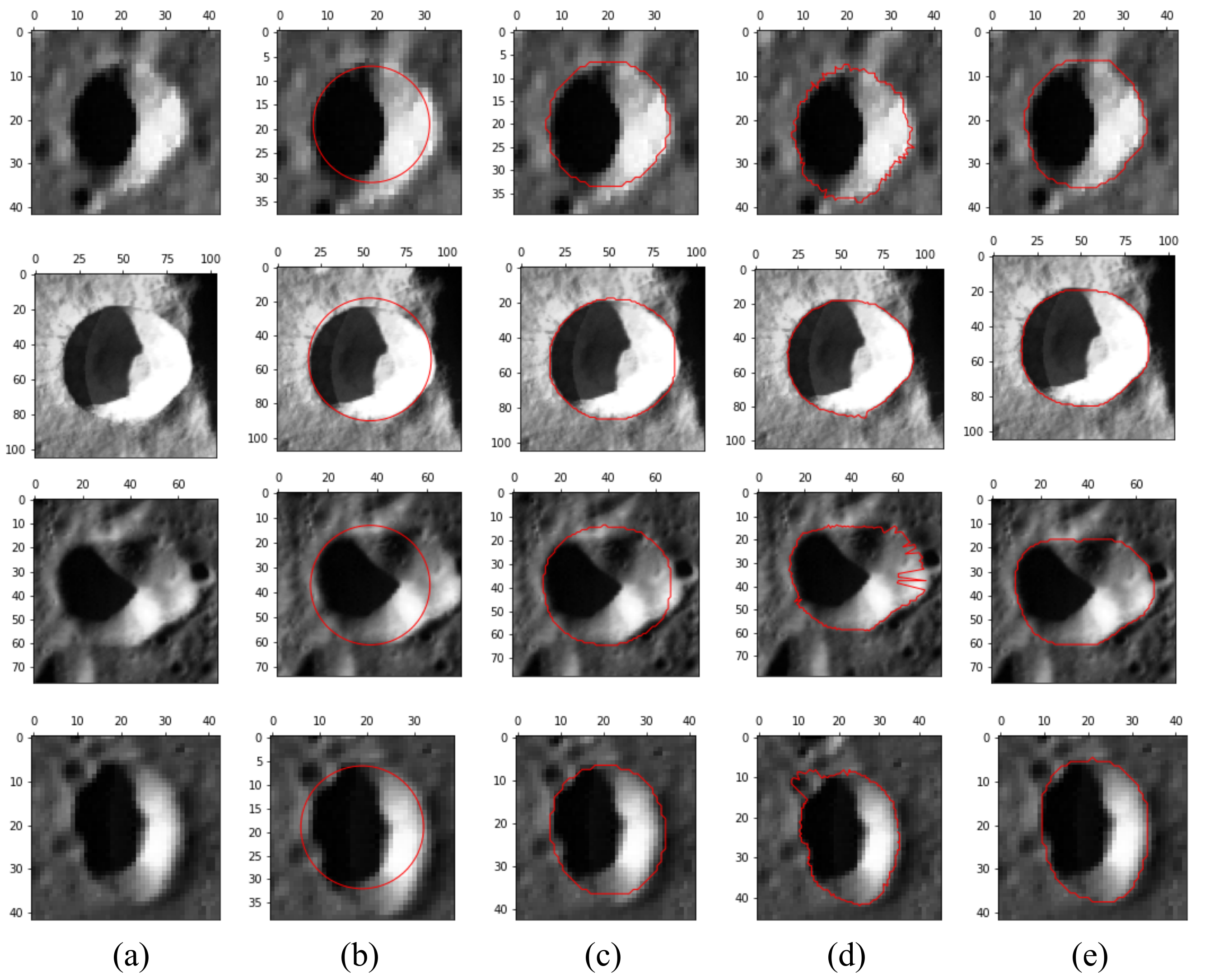}
	\caption{Visualization of craters' shapes in optical-images (crater rim in red color). (a) Original image, Crater rim marked using different methods: (b) Catalog (Povilaitis et al.~\cite{povilaitis2018crater}), (c) Mask R-CNN trained with circular masks, (d)  Proposed adaptive rim estimation method, (e) Proposed adaptive rim estimation method with semi-supervised deep learning.}
	\label{fig:csr_explain}
\end{figure*}

Figure~\ref{fig:csr_explain} shows a few sample craters for visual inspection of the extracted shapes. 
The extracted shapes in the proposed approach (Figure~\ref{fig:csr_explain}(d), Figure~\ref{fig:csr_explain}(e)) are more accurate as compared to the circular shapes (Figure~\ref{fig:csr_explain}(b)) and the deep learning framework, i.e., Mask R-CNN with training circular mask (Figure~\ref{fig:csr_explain}(c)). 
Also extracted rims from the proposed adaptive rim estimation (Figure~\ref{fig:csr_explain}(d)) are further smoothed out after applying semi-supervised deep learning approach (Figure~\ref{fig:csr_explain}(e)).
As evident from Figure~\ref{fig:csr_explain}, the proposed systems extract more accurate shapes of the craters as compared to a circular mask. 
In the absence of a ground truth catalog with manually marked rims, a more detailed quantitative measure of improvement is not feasible and will be addressed as part of future work.  
The extracted shapes will be shared with the published version of the paper.

\textbf{\textit{User Study:}}

Similar to Qi  et al.~\cite{qi2021embedding}, we also conducted a user study for the evaluation of crater shapes. 
We asked users to give rank to the results of different methods based on the rim quality. 
Following are the four methods chosen for the user study:
\begin{itemize}
    \item \textbf{circular}: Crater rims marked as circles using Povilaitis et al. catalog~\cite{povilaitis2018crater},
    \item \textbf{mrcnn\_circular}: Mask R-CNN trained with circular mask,
    \item \textbf{proposed\_arm}: Proposed adaptive rim estimation method,
    \item \textbf{proposed\_armdl}: Proposed adaptive rim estimation method with semi-supervised deep learning.
\end{itemize}

\begin{figure*}[!htb]
    \centering
    \includegraphics[width=0.97\textwidth,trim=0.18cm 0.18cm 0.18cm 0.18cm, clip]{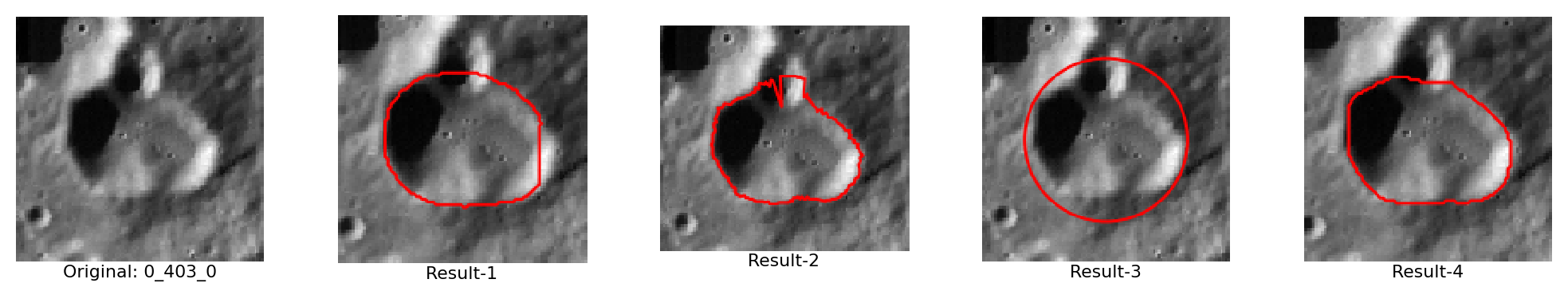}
    \caption{Example of images provided to the users for ranking different methods. The rims are highlighted in red. Method that followed closest to the actual boundary will have the best rank, i.e., 1.}
    \label{fig:eg_ranking}
\end{figure*}

The ranking order is 1 to 4; 1 indicates that the crater has the best rim quality, whereas 4 indicates the lowest rim quality.
As extraction of exact shapes is more beneficial for non-circular craters, we have considered only those craters which have sphericity less than $0.7$. 
Also, users may find it difficult to visually inspect small craters; therefore, we have considered craters of diameter $\geq10$ km.  
Out of all the craters satisfying these two constraints, we have randomly selected 200 craters from the test region for user evaluation. 
For each crater, we have provided an original image of the crater and the results of four methods, where boundaries are highlighted in red (e.g., Figure~\ref{fig:eg_ranking}). 
The order in which these 200 craters are shown to the users and the ordering of results for each crater is random to avoid any biases. 
For example, one possible ordering will be: Result-1: mrcnn\_circular, Result-2: proposed\_armdl, Result-3: circular, Result-4: proposed\_arm. The users only see Result-1, Result-2, Result-3 and Result-4 as captions of the results' images and do not see the name of the methods. 

\begin{table*}[ht!]
    \centering
    \caption{Results of the user study for comparison of crater shapes estimated using different methods. Ranking order: 1 indicates the best rim quality and 4 is the lowest.}
    \label{tab:ranking}    
\begin{tabular}{p{0.01\textwidth}p{0.03\textwidth}|p{0.2\textwidth}p{0.2\textwidth}p{0.2\textwidth}p{0.2\textwidth}}
 \multicolumn{4}{c}{\hspace{8cm} \textbf{Methods}} \\
     & & circular (\%)   &  mrcnn\_circular (\%)   & proposed\_arm (\%)    & proposed\_armdl (\%)   \\ \cline{2-6}
\multirow{4}{*}{\rotatebox[origin=c]{90}{\textbf{Ranking}}}
&  1  & 2.70 &  9.67 &  24.10 &  63.53 \\
 & 2  & 5.70 &  34.8 &  69.58 &  89.92 \\
  & 3  & 14.12 &  96.30 &  92.20 &  97.38 \\
&   4 & 100 &  100 &  100 &  100 \\
\end{tabular}
\end{table*}

This user study involved 30 users. 
We also collected statistics of some of the characteristics related to the users, such as age and background in computer science. 
Age of users (in years): \{‘16-25’: 40\%, ‘26-35’: 53.33\%, and ‘36-60’: 6.67\%\}, gender of users: \{‘Male’: 83.33\%, ‘Female’: 16.67\%\}, background in computer vision/image processing/deep learning/remote sensing: \{‘No course’: 30\%, ‘1 course’: 23.33\%, ‘2 courses’: 16.67\%, ‘3 courses’: 13.33\%, ‘4 courses or more’: 16.67\%\}. 

We have 30 users and 200 craters, so the total samples for evaluation are 6000. 
Out of these, 3812 times proposed\_armdl and 1446 times proposed\_arm have the best rank, i.e., 1.
Table~\ref{tab:ranking} shows the cumulative ranking of craters' shapes for each method.
Our proposed adaptive rim estimation method with semi-supervised deep learning (proposed\_armdl) has the highest percentage of best rank (63.53\%). 
Further, in the top 2 ranks, our proposed\_armdl has the highest percentage, i.e., 89.92\%. 
Of 30 users, for 26 users, the best ranked method is proposed\_armdl for the largest number of craters, out of the 200 craters. 
And for the remaining 4 users, the best ranked method is proposed\_arm for the largest number of craters, out of the 200 craters. 
Hence, it can be concluded that our proposed\_armdl method has the best crater shape performance.

\subsection{Transfer learning on Martian Surface}
\label{subsec:Transfer learning on Martian Surface}
In order to show the robustness of the proposed method, we have evaluated our method on the Martian surface. We utilized a dataset generated by Tewari et al.~\cite{TEWARI2022105500} in our work, and Martian Robbins catalog~\cite{robbins2012new} is used for performance evaluation.
Similar to Tewari et al.~\cite{TEWARI2022105500}, we have done fine-tuning in the following manner: the weight and biases of the head layer of Mask R-CNN are unfrozen, whereas other weights of Mask R-CNN are kept frozen. We achieved precision 62.05\%, recall 79.59\%, $F_1$-score 69.73\%, and $F_2$-score 75.33\%. 

It can be observed that the proposed method is able to detect more than 75\% of the crater from the ground truth.

\section{Conclusion and Future Work}
\label{sec:Conclusion and Future Work}
This paper presented a system for accurately extracting the shapes of craters. 
It mainly consisted of two components. 
First, an unsupervised non-deep learning approach using elevation profile and morphological operations is used to extract the shapes of the craters. 
Second, the extracted shapes are further refined using semi-supervised deep learning. 
The proposed system can detect almost all craters from the catalog and also detect new craters. 
Further, we calculated morphological features that can be used to understand craters' characteristics and help differentiate other planetary features. 
Through visual inspection, we found that most of the extracted shapes are accurately extracted. 
However, due to the non-availability of ground truth shapes, quantitative verification is not done in this work. 
Future work will include creating a ground truth dataset for quantitative verification and further improving crater shape retrieval.

\section*{Acknowledgment}
This study is based upon the work partially supported by the ISRO, Department of Space, Government of India under the Award number ISRO/SSPO/Ch-1/2016-17.  
Atal Tewari is supported by TCS Research Scholarship. 
We are grateful to Amrita Singh and K Prateek, students in MANAS Lab, for discussion on morphological features of the craters. 
Any opinions, findings, conclusions, or recommendations expressed in this material are solely those of the author(s) and do not necessarily reflect the funding agencies' views.
We also acknowledge the use of data from NASA's LRO spacecraft, which was downloaded from the archives of the USGS.  

\bibliographystyle{IEEEtran}
\bibliography{IEEEabrv,paper_bib}
\end{document}